# Quantum scattering engineering for the reduction of dark current in very long wavelength quantum well infrared photodetector.

Emmanuel Lhuillier[1,2], Emmanuel Rosencher[1], Isabelle Ribet-Mohamed[1], Alexandru Nedelcu[3], Laetitia Doyennette[2], Vincent Berger[2].

[1] ONERA, Chemin de la Hunière, 91761 Palaiseau cedex, France.
[2] Matériaux et Phénomènes Quantiques, Université Paris 7, Bat. Condorcet, Case 7021, 75205 Paris cedex 13, France.
[3] Alcatel-Thales III-V Lab, Campus de l'Ecole Polytechnique, 1 Avenue A. Fresnel, 91761 Palaiseau cedex, France.

**Abstract**
Dark current is shown to be significantly reduced in quantum well infrared photodetectors in the tunneling regime, i.e. at very low temperature, by shifting the dopant impurity layers away from the central part of the wells. This result confirms that the interwell tunneling current is dominated by charged impurity scattering in usual structures. The experimental results are in good quantitative agreement with the proposed theory. This dark current reduction is pushing further the ultimate performances of quantum well infrared photodetectors for the detection of low infrared photon fluxes. Routes to further improvements are briefly sketched.

Quantum well infrared photodetectors (QWIP) have been extensively used[1,2] for the detection of low infrared photon flux which is of utmost importance in aerospace applications for instance. The operating temperature for QWIP is generally in the 50 to 75K range[3] for 10µm detectors and in the 35 to 60K range[3,4] around 15µm. These temperatures are a compromise between the level of performance of the detector and the lifetime of the cryogenic cooling device. It has already been demonstrated that for QWIP operating in the VLWIR (very long wavelength infrared), the dark current level is the point on which performance improvements have to be focused. An increase of performance can generally be reached by decreasing the detector temperature. Such an improvement is possible as long as the dark current is dominated by thermoionic emission[5]. However, at sufficiently low temperature, the magnitude of the dark current is driven by the residual tunnel coupling between two following wells. The detector performances become independent of the temperature so that the performance improvements require a structure optimization.

In a previous paper we have suggested that for VLWIR QWIP operating in the tunnelling regime, the dark current mostly results from the interaction between the electron and the doping ionized impurities[6]. A change in the QWIP doping profile may thus allow a decrease of the dark current. Usually the doping is located in the central part of the well in QWIP. Changing the doping profile has already been proposed in the literature but for different purposes. In order to solve doping segregation problems, Schneider et al[7,8] have proposed to move the doping away from its central position to the first part of the well. Luna et al[9,10] suggested to design modulation doped QWIP in order to improve their responsivity. The effect of the doping position on the spectral response has also been studied by Dupont et al[11] for the control of the transition linewidth and by Pan et al[12] for the possibility to observe forbidden transitions. In this paper, we propose to investigate the influence of the doping position on the magnitude of the dark current. Structures where dark current is divided by a factor of ≈ 2, *mutatis mutandis*, will be presented.

In order to predict quantitatively the effect of the doping position we developed a hopping transport model which includes interaction of the electrons with LO phonon, LA phonon, alloy disorder, interface roughness and ionized impurities. Wave functions have been calculated in a two wells structure using a two bands k·p method[13] and self consistent Poisson/Schrödinger code[6].

In long wavelength QWIPs and for moderate electric field, we have demonstrated that the electron ionized interaction is the one which drives the dark current[6] in the tunnel regime. Dark current reduction may thus result from a reduction of the impurity mediated scattering rate between the ground states of two adjacent wells. It is interesting to consider the ionized impurities scattering rate expression[14,15]:

$$\Gamma_{II} = \frac{e^4}{8\pi\hbar\varepsilon_0^2\varepsilon_r^2}\frac{m^*}{\hbar^2}\int_{impurities}dz_i N(z_i) \times \int d\theta \frac{F_{II}(|K_i - K_f|)}{|K_i - K_f|^2}$$

where e is the proton charge, m* the GaAs effective mass, $\hbar$ the reduced Planck constant, $\varepsilon_r$ the GaAs permittivity, $z_i$ the impurity position, $N(z_i)$ the volumic doping profile, $K_i$ and $K_f$ the initial and final wavevectors and $\theta$ the angle between the two vectors $K_i$ and $K_f$. Finally the overlap integral $F_{II}(Q) = \left| \int dz \xi_f^*(z) e^{-Q|z-z_i|} \xi_i(z) \right|^2$ is the form factor of this interaction, which links the geometry of the device to the magnitude of the scattering: $\xi_i$ (resp. $\xi_f$) is the initial (resp. final) electron envelope wavefunction.

A careful examination of the form factor expression immediately indicates that, because of the overlap integral between $\xi_i$ and $\xi_f$ mediated by the $e^{-Q|z-z_i|}$ term, moving the doping away from the central part of the well will reduce the form factor and the associated dark current. A possible solution would be to localize the doping impurities in the barriers. However, this is liable to i) introduce deep levels into the barrier[16] which will be detrimental to the dark current level and ii) create quantum levels inside the barrier[17,18], which is also detrimental to transport properties. We have thus chosen to move the crenel of doping away from a central position of the well to the border of the well. The shift has been chosen towards the surface of the sample rather than towards the substrate in order to avoid a cancelation of the expected effect by doping segregation problems.

tab. I : Measurements of well width, barrier width and aluminium content using X ray diffraction

| Device | A | B |
|---|---|---|
| Doping position | Central doping | Shifted doping |
| Aluminium content (%) | 15.6±0.1 | 15.5±0.1 |
| Well width (nm) | 6.7±0.1 | 6.7±0.1 |
| Barrier width (nm) | 39.2±0.1 | 38.9±0.1 |

We have designed two structures which are expected to differ only by their doping profile. The structure is a forty periods QWIP grown by molecular beam epitaxy. The nominal well width is 6.8 nm, whereas the barrier width is 39 nm. The aluminium content of the barrier is 15.5%. This results in a peak transition around 13.5µm. The doping sheet density is the same for the two devices and equals $3\times10^{11}\text{cm}^{-2}$. Structure A (reference) is doped in its central third whereas structure B is doped in its last third (surface side), see FIG. 1. Precise measurements of the well width, barrier width and aluminium content have been obtained using X ray diffraction and results are presented in tab. I.

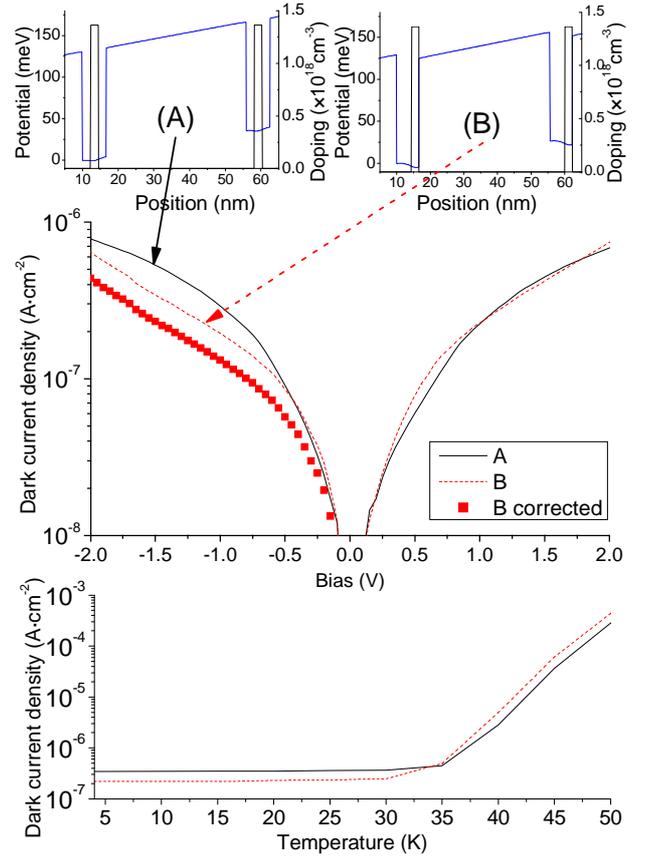

FIG. 1 Upper part: self consistent calculation of the potential profile under an electric field of 8kV·cm$^{-1}$. Central part: Dark current measurements in the tunneling regime as a function of the applied bias for device A and B, T=4K. The dot curve shows the corrected current for structure B. Lower part: Current as a function of the temperature for the two devices, under a -1.1V bias.

Those samples have been processed into mesas of 50 and 100µm. The resulting devices are mounted on the cold finger of an helium cryostat. The temperature is regulated with a Lakeshore 331 thermal controller. The current is measured with a sub-femtoampermeter (Keithley 6430). Spectral measurements have been realized with a Bruker Equinox 55. Quantum efficiency measurements under low infrared flux have been obtained using a double cryostat device: The first cryostat is used to cool the detector while the second cryostat, operated with nitrogen cools down the blackbody. The numerical aperture of the system is f/2.8.

I(V) measurements for the two devices are given on FIG. 1 (central part) in the tunneling regime (T=4 K). As expected the dark current is reduced for the device with the shifted doping for negative bias. In this case the electric field tends to localize the wave function at the opposite of the doping, decreasing the overlap integral. On the contrary for positive bias, the electric field moves the electron wave function closer to the doping which tends to inverse the effect. FIG. 1 (lower part) shows the dark current as a function of the temperature. At high temperature the reference device presents the lowest current. This clearly results from the difference of confinement of the electron. Indeed spectral measurements, presented on FIG. 2, show a lower peak energy for the reference device which indicates a higher confinement. Such a result is confirmed by X-Ray measurements, see tab. I, since the reference presents larger and higher barrier. Due to this composition fluctuation the activation energy is smaller for the reference which leads to a reduced dark current for this device when it operates in its thermionic regime. At low temperature, the B device is the one with the smaller dark current, in spite of this lower confinement. The dark current reduction is of 30% in the -1.5V → -1V range of bias, which is quite close to the expected decrease. To have an idea of the current which we may have obtained in the case where the two samples only differed by their doping position, we plot, on FIG. 1, the experimental current multiplied by the ratio of the tunnelling probability for structures A and B, thus we expect to have corrected the effect linked to the difference in the barrier size. Its expression is given by:

$$J_{corrected} = J_{experimetal} \times \frac{P(E_1(A))}{P(E_1(B))}$$

where $E_1$ is the fundamental level energy and P is given by the WKB approximation[19]

$$P(E) = \exp\left(-\frac{4\sqrt{2m_b^*}}{3F\hbar}\left[(Vb-E)^{3/2} - (Vb-eFL_b-E)^{3/2}\right]\right)$$

Thus the corrected effect of the shift of the doping is a 50% decrease of the dark current.

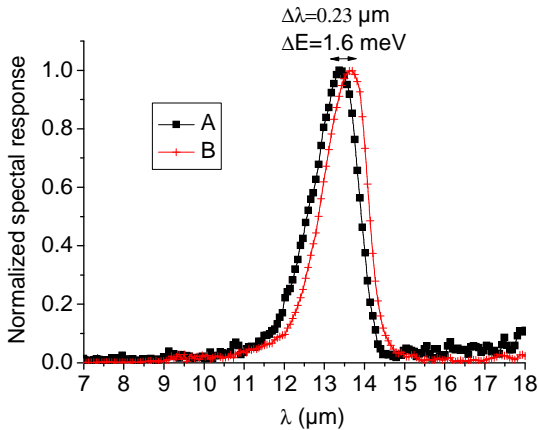

FIG. 2 Spectral measurements for devices A and B, under -1V. The period of the coupling grating is 4.2µm.

Finally using 300K absorption measurements we have checked that the absorption is similar for the two devices, which means that the doping levels are very close. Thus we do not expect that the dark current decrease results from a change in the doping level. Thus, in spite of the lower confinement, the optimized device presents a reduced dark current.

Using the parameters obtained by X-Ray diffraction we can compare the theoretical interwell scattering rates with the experimental datas. The experimental interwell scattering rates are obtained from the expression $\Gamma = \frac{J}{e \cdot n_{2D}}$ where J is the current density and $n_{2D}$ the sheet carrier density. We also assume that the electric field is constant over the whole structure. We obtained a reasonable agreement for the dark current reduction value between theory and experimental data. The difference of the shape of the curves may result from electric field inhomogeneities.

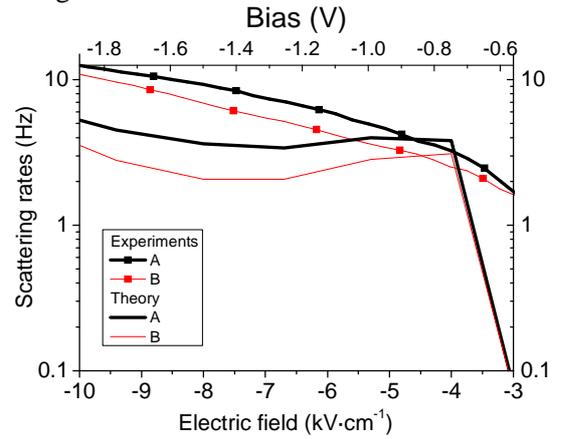

FIG. 3: Experimental and theoretical scattering rates as a function of the applied bias. The theoretical current has been obtained for a doping of $n_{2D}=5\times10^{11}$cm$^{-2}$, for a better agreement of the scattering rate magnitude.

Finally, the external quantum efficiency is almost the same for the two devices: 13.6% for device A and 11.5% for device B, under -2V.

It is possible to further increase this reduction of the dark current. Indeed, for the B structure, the shape of the energy band profile is affected by the electrostatic reconfiguration. A self consistent evaluation of the energy band profile (EBP) is shown FIG. 1. Keeping all growth parameters constant (well and barrier width, aluminium content) the change of the EBP, due to the shift of the doping position, increases the overlap $\langle\Psi_n|\Psi_{n+1}\rangle$ between the ground states by a factor three (with $\Psi_n$ is the ground state wave function of the n$^{th}$ well). The matrix element associated with ionized impurities $\langle\Psi_n|\frac{e^2}{4\pi\varepsilon_0\varepsilon_r}\frac{1}{r}|\Psi_{n+1}\rangle$ is reduced but at the same time the matrix element associated with other interactions raise. Consequently it will be much more favourable to build a symmetric doping profile. For this we can split the crenel of doping in two smaller crenels, each one being located on the edge of the well. However such a sample may be limited by the doping segregation.

To conclude we have proposed and tested an alternative way to the barrier width increase for the reduction of the dark current in the tunnelling regime. This technique is based on the quantum scattering engineering of the interwell scattering rate. This results in the minimization of the scattering overlap integral between electron states in adjacent QWs by a shift of the doping position towards the border of the QWs. This method allows a reduction of 50 % of the tunnelling current while keeping the quantum efficiency almost unchanged.